\documentclass[sigconf, nonacm]{acmart}
\AtBeginDocument{%
  }

\setcopyright{none}
\copyrightyear{2026}
\acmYear{2026}

\usepackage{listings}

\usepackage{xcolor}
\usepackage{float}
\usepackage{caption}
\usepackage{changepage}
\usepackage{multirow}

\graphicspath{{./images/}}

\definecolor{codegreen}{rgb}{0,0.6,0}
\definecolor{codegray}{rgb}{0.5,0.5,0.5}
\definecolor{codepurple}{rgb}{0.58,0,0.82}
\definecolor{backcolour}{rgb}{0.95,0.95,0.92}

\lstdefinestyle{mystyle}{
    backgroundcolor=\color{backcolour},   
    commentstyle=\color{codegreen},
    keywordstyle=\color{magenta},
    numberstyle=\tiny\color{codegray},
    stringstyle=\color{codepurple},
    basicstyle=\ttfamily\footnotesize,
    breaklines=true,
    numbers=left,
    numbersep=5pt,
    showstringspaces=false,
    tabsize=2,
    aboveskip=0.1em,
    belowskip=0.1em,
}
\lstdefinelanguage{CalyxPseudo}{
    morekeywords=[1]{par, seq, if, execute},
    morekeywords=[2]{int, for},
    keywordstyle=[1]\color{blue},
    keywordstyle=[2]\color{magenta},
    morecomment=[l]{//},
    morestring=[b]",
    sensitive=true,
    alsoletter={_, -},
}

\acmConference[C4ML’26]{7th Compilers for ML Workshop (C4ML 2026)}{Sunday, Feb 1, 2026}{Sydney, Australia}

\begin{document}

\title {From PyTorch to Calyx: An Open-Source Compiler Toolchain for ML Accelerators}
 
\author{Jiahan Xie}
\email{jxie84@ucsc.edu}
\affiliation{%
  \institution{University of California, Santa Cruz}
  \city{Santa Cruz}
  \state{California}
  \country{USA}
}

\author{Evan Williams}
\email{emw236@cornell.edu}
\affiliation{%
  \institution{Cornell University}
  \city{Ithaca}
  \country{New York}
  \country{USA}
}

\author{Adrian Sampson}
\email{asampson@cs.cornell.edu}
\affiliation{%
  \institution{Cornell University}
  \city{Ithaca}
  \country{New York}
  \country{USA}
}

\renewcommand{\shortauthors}{Xie et al.}

\begin{abstract}
We present an end-to-end open-source compiler toolchain that targets synthesizable SystemVerilog from ML models written in PyTorch. Our toolchain leverages the accelerator design language Allo, the hardware intermediate representation (IR) Calyx, and the CIRCT project under LLVM. We also implement a set of compiler passes for memory partitioning, enabling effective parallelism in memory-intensive ML workloads. Experimental results demonstrate that our compiler can effectively generate optimized FPGA-implementable hardware designs that perform reasonably well against closed-source industry-grade tools such as Vitis HLS. 
\end{abstract}

\maketitle

\section{Introduction}
High-level synthesis (HLS) is a promising approach for designing accelerators for ML applications \cite{hls4ml, fpgaconvnet, DnnWeaver, Dahlia}. HLS allows us to compile high-level functional specifications from software (typically written in C or C++) to synthesizable RTL suitable for hardware implementation. However, few HLS approaches exist for compiling directly from the languages and frameworks that express ML models, such as PyTorch. Moreover, the prior work relies heavily on closed-source commercial HLS toolchains, such as Xilinx Vitis HLS. Recent MLIR-based HLS systems demonstrate the potential for open-source compiler stacks for accelerator design \cite{ScaleHLS}. 

Building such a system introduces several challenges. ML models are typically written in Python and rely heavily on floating-point arithmetic, complex tensor operators, and multi-module execution patterns, none of which map cleanly onto low-level hardware descriptions. Supporting these workloads requires a compilation flow that can translate high-level tensor programs into hardware-oriented IRs while preserving enough structure to enable parallel execution and aggressive memory optimizations. However, traditional HLS approaches are limited, as high-level software programming languages are not well-suited for direct translation to hardware semantics. Additionally, modern ML accelerators must exploit fine-grained parallelism, but parallel execution introduces hazards such as memory port contention that require careful program analysis. 



Therefore, we introduce an end-to-end open-source compiler toolchain that generates synthesizable SystemVerilog from PyTorch models through a structured compilation flow. Our system uses Allo \cite{Allo} to translate PyTorch programs into MLIR, leverages domain-specific MLIR dialects \cite{UrbachPetersen22} to preserve high-level tensor structure, and relies on the CIRCT \cite{CIRCT,DemmeLandy22} infrastructure to perform hardware lowering. The resulting program is expressed in Calyx, whose explicit separation of control and hardware structure allows us to encode accelerator architectures and optimizations cleanly. Finally, we compile Calyx to SystemVerilog and use standard FPGA vendor tools such as Xilinx Vivado to synthesize, place–and–route, and deploy accelerators. Our contributions are as follows:
\begin{itemize}
    \item An end-to-end open-source compiler stack from PyTorch to synthesizable SystemVerilog using Allo, CIRCT, and Calyx.
    \item Memory banking and partitioning analyses in Calyx enabling safe and efficient parallel access patterns.
    \item FPGA evaluation showing performance gain up to $2.21\times$ Vitis HLS.
\end{itemize}

\section{Background}
\subsection{Allo}
Allo \cite{Allo} is a compiler for constructing large-scale, high-performance hardware accelerators. Its lowering pipeline translates PyTorch programs into MLIR while preserving tensor semantics, control flow, and data-layout information. The resulting structured MLIR program integrates cleanly with downstream dialects and compiler infrastructures, forming a bridge between high-level ML frameworks and hardware generation backends.

\subsection{Calyx}
Calyx \cite{calyx} is an IR and compiler infrastructure designed for generating hardware accelerators from high-level programming languages. It explicitly separates control flow from structural hardware descriptions, enabling optimizations that leverage both perspectives. The Calyx compiler then lowers the program into synthesizable RTL through a series of transformation and optimization passes. 


\subsection{CIRCT}
Circuit IR Compilers and Tools (CIRCT) \cite{CIRCT} is an open-source, LLVM-based infrastructure for building hardware compilers. Built atop MLIR \cite{MLIR}, CIRCT provides a suite of hardware-specific dialects and transformation passes to support the development of custom compilation flows, hardware synthesis pipelines, and IRs.

Calyx is integrated into CIRCT as one of its dialects and is particularly valuable because CIRCT provides an MLIR-native path down to a hardware-oriented dialect, and Calyx is the natural endpoint of that lowering because it retains both structural hardware description and software-style control, giving us a unified space to express accelerator-specific rewrites before RTL generation. 


\section{Contribution}
\subsection{Overview}
The goal of this work is to run PyTorch programs on custom hardware accelerators. We use FPGAs as the prototyping platform and develop a fully open-source compilation pipeline that transforms high-level Python programs into synthesizable hardware designs.

Our toolchain is composed of several open-source components. We begin by using the Allo framework~\cite{Allo} to compile PyTorch models into MLIR programs. These MLIR programs are then progressively lowered via native MLIR passes to dialects supported by CIRCT. During this stage, we also apply high-level optimizations to improve performance and hardware compatibility.

Once the program reaches a form compatible with Calyx, we emit code in the Calyx IR. The Calyx compiler then performs hardware-specific transformations and generates synthesizable SystemVerilog. This hardware design is finally deployed to an FPGA, completing the software-to-hardware compilation path.

\subsection{Lowering to Calyx}
Figure 1 shows the compilation flow we developed and orchestrated to lower PyTorch to Calyx. We leverage Allo to lower to the MLIR Linalg dialect. We then lower to the MLIR Affine dialect, and then the Memref and SCF dialects. We use CIRCT to produce Calyx, and Vivado HLS to execute the design on an FPGA. 

Building upon \cite{UrbachPetersen22}, which developed initial lowering support for structured control flow (SCF) constructs (e.g., \texttt{for}, \texttt{while}), we expanded this work by supporting additional SCF operations, including \texttt{if} and \texttt{parallel}. We also implemented the translation of software-level functions into Calyx components: each function becomes a hardware module with ports for scalar arguments and instantiated memories for tensor arguments, while function calls become component instantiations. We developed a full, general floating-point library for Calyx to provide floating-point support in CIRCT, including the  integration of the Berkeley HardFloat \cite{HardFloat} components. We also implemented bit-level IEEE-754 constant handling and human-readable attribute representations to enhance visibility and debuggability in CIRCT.

Collectively, these contributions form the foundation of our compiler stack, enabling high-level MLIR programs to be lowered into the Calyx IR and synthesized into RTL.

\begin{figure}[!t]
  \centering
  \includegraphics[height=2.6in]{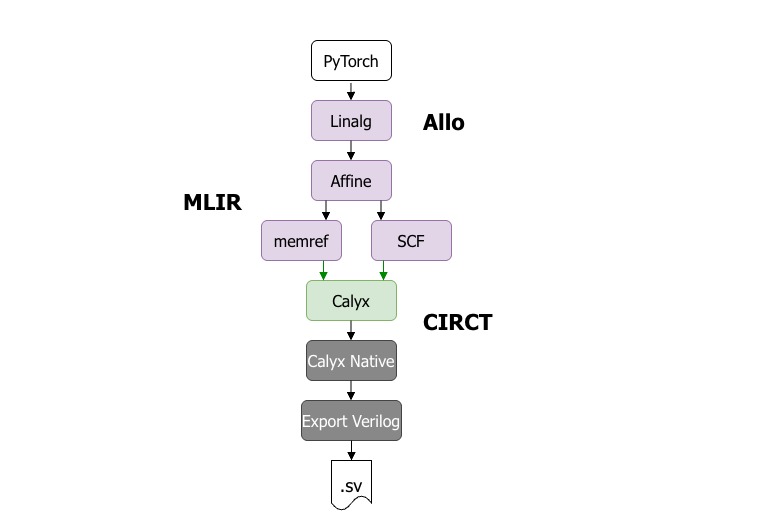}
  \caption{Compilation pipeline from PyTorch through Allo to Calyx.}
  \label{fig:allo-calyx}
\end{figure}

\subsection{Memory and Parallelism Optimizations}
To optimize the forward pass of the model, we leverage parallelism to increase hardware throughput. Calyx supports parallel execution as a first-class control construct, allowing us to explicitly model concurrent computation. Our task, then, is to expose and maximize parallelism - particularly in memory access patterns - while adhering to hardware constraints.

We base our approach on data parallelism: we partition memories into multiple banks so that different operations can access data concurrently. Since Calyx assumes each memory supports only a single read or write per cycle, banking allows us to duplicate storage and route accesses to different banks.

Our data-parallel strategy must avoid bank conflicts: memory banking increases the number of ports and introduces control logic to select which bank to access, often resulting in nested conditionals. If not optimized, these additional control paths increase latency and resource usage.

Our implementation supports cyclic memory partitioning, and we assume this scheme throughout. To route accesses to the correct bank, we use a \texttt{switch} statement (or nested \texttt{if}-\texttt{else} chains when \texttt{switch} is unavailable). A naive implementation that directly emits control branches for each bank leads to code-size blow-up. For a memory with $d$ dimensions and a partition factor $c$, the number of unique control branches scales as $c^d$. In Calyx, these branches are all instantiated as hardware, even if only one is active at runtime. This not only increases area usage but also results in deeper control finite state machines (FSMs), which hurt performance.


To illustrate the problem concretely, consider the following simple loop writing to a four-element memory with a memory banking factor of 2:


\begin{lstlisting}[language=CalyxPseudo]
for (int i = 0; i < 4; ++i) {
    if (i % 2 == 0) {
        mem_bank_0[i / 2] = i;
    } else {
        mem_bank_1[i / 2] = i;
    }
}
\end{lstlisting}

To parallelize this loop, we materialize it with a factor of 2 using nested \texttt{seq} and \texttt{par}:

\begin{lstlisting}[language=CalyxPseudo]
seq for (int i = 0; i < 2; ++i) {
    par for (int j = 0; j < 2; ++j) {
        int new_index = 2 * i + j;
        if (new_index % 2 == 0) {
            mem_bank_0[new_index / 2] = new_index;
        } else {
            mem_bank_1[new_index / 2] = new_index;
        }
    }
}
\end{lstlisting}


However, each \texttt{par} block must be unrolled into separate arms with statically known indices. In our example, this produces two parallel arms—\texttt{j = 0} and \texttt{j = 1}—each computing a distinct \texttt{new\_index}. Although each arm activates only one side of the banking condition at runtime, Calyx still instantiates both sides of the \texttt{if}–\texttt{else} because the predicate cannot be folded symbolically. As a result, even conflict-free access patterns may lead to multiple arms writing to the same physical bank, creating write hazards that Calyx cannot eliminate statically.

We implement two techniques to ensure safe and efficient memory parallelism: (1) we express banking by embedding the bank index into the memory’s dimensional layout instead of guarding accesses with conditional logic; and (2) we rewrite loop nests whose structure would otherwise duplicate sequential controllers when executed in parallel.

For the first, rather than emitting branch logic per access, we raise the memory’s dimensionality and bake the bank index into the first dimension. For instance:

\begin{lstlisting}[language=CalyxPseudo]
seq for (int i = 0; i < 2; ++i) {
    par for (int k = 0; k < 2; ++k) {
        mem[k][i] = 2 * i + k;
    }
}
\end{lstlisting}

Unrolling this loop yields:
\begin{lstlisting}[language=CalyxPseudo]
seq for (int i = 0; i < 2; ++i) {
    parallel execution {
        execute par-arm-0 {
            mem[0][i] = 2 * i + 0;
        }
        execute par-arm-1 {
            mem[1][i] = 2 * i + 1;
        }
    }
}
\end{lstlisting}


Here, the bank index is a compile-time constant in each parallel arm, ensuring that memory accesses are disjoint and contention-free. This approach enables us to preserve parallelism without incurring the cost of unnecessary control overhead.

The second transformation operates at the level of loop structure. This example also highlights a broader point: loop transformations that are semantically equivalent in software do not necessarily yield equivalent hardware. Consider two nestings: one where \texttt{seq(i)} surrounds \texttt{par(j)}, and another where \texttt{par(j)} surrounds \texttt{seq(i)}. Although they are semantically equivalent to \texttt{par(j)} around \texttt{seq(i)} in software, they behave very differently in hardware. In the first form, there is a single sequential controller that iterates over \texttt{i}, and each iteration triggers a parallel group over \texttt{j}. In the second form, however, each parallel arm receives its own private sequential controller for iterating over \texttt{i}, effectively duplicating the entire FSM. This replication inflates the hardware area and increases control overhead. To prevent such unnecessary duplication, our compiler detects these patterns and rewrites parallel–sequential loop nests into schedules that share control logic while preserving the intended parallelism.

Through these memory and loop-aware transformations, our compiler produces parallel Calyx programs that are both correct and efficient for hardware execution.

\section{Results and Evaluation}
In this section, we evaluate our Calyx-based flow against Vitis HLS under two configurations. In Section~\ref{sec:baseline}, we compare baseline designs with no data parallelism: neither toolchain applies any banking strategy. In Section~\ref{sec:banked}, we enable memory banking for both flows using matched partitioning factors, schemes, and dimensions, allowing a direct comparison of parallelized configurations.

Before presenting results, we note that Vitis HLS incorporates many mature and often implicit optimizations that are not currently implemented in our Calyx-based flow and cannot be disabled or inspected through \texttt{pragma}s. Thus, we do not expect Calyx to outperform Vitis in baseline latency or resource usage. Instead, our goal is to assess how competitive Calyx can be and whether our targeted banking transformations meaningfully narrow the gap. As shown below, while Vitis maintains an advantage in baseline configurations, Calyx becomes competitive once banking is enabled, offering an open-source and compiler-controlled foundation for further optimization.

\subsection{Benchmark Models}
We evaluate the performance and resource usage of our compiler by benchmarking three representative ML models and comparing them against a commercial HLS toolchain, Xilinx Vitis HLS. The models include a feed-forward neural network (FFNN), a convolutional neural network (CNN), and a multi-head attention (MHA) module.

The FFNN model takes an input of 64 features, followed by a fully connected layer of size $64 \times 48$, a ReLU activation, and a second fully connected layer of size $48 \times 4$.

The CNN model processes a $80 \times 60$ color image with 3 channels. The first layer performs a 2D convolution with $5 \times 5$ kernels, 3 input channels, and 8 output channels using unit strides. This is followed by a ReLU activation and a max-pooling layer with a $2 \times 3$ window. The resulting feature map is flattened and passed through a fully connected layer for binary classification.

The MHA model is derived from the Transformer architecture and uses 2 attention heads. Each head operates on a 21-dimensional subspace of a 42-dimensional embedding, with causal masking for autoregressive decoding.

\subsection{Comparison Across Models}
\label{sec:baseline}
For comparison, we use Vitis HLS, a widely used commercial HLS tool. The Allo project provides a shared frontend that lowers PyTorch models to MLIR, which is then further compiled to either HLS C++ for Vitis or Calyx for our flow. To ensure fairness, the MLIR input is held constant across both paths. All HLS pragmas are disabled except for \texttt{\#pragma ARRAY\_PARTITION}, which is used to compare against our memory banking configuration. We ensure consistent banking factors, schemes (cyclic), and dimensions across both flows.

Wall-clock latency is computed from each design’s cycle count and its maximum post–place-and-route frequency that meets timing, obtained by sweeping the timing constraint and selecting the implementation that achieves the lowest overall latency.

\begin{figure}[t]
  \centering
  \includegraphics[width=0.9\columnwidth]{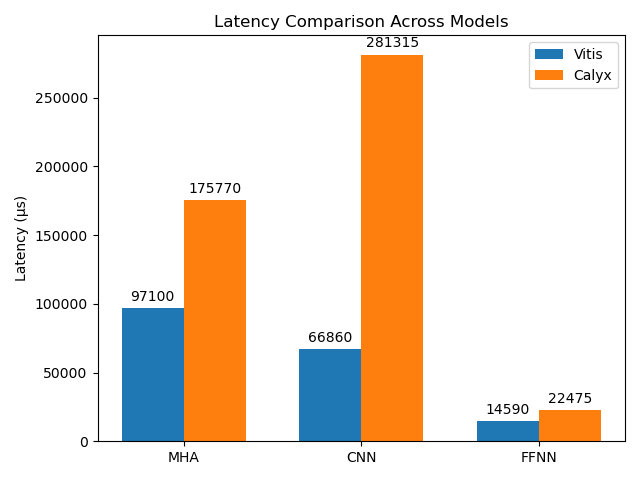}
  \caption{Wall-clock latency comparison across models.}
  \label{fig:models_latency}
\end{figure}

Figure~\ref{fig:models_latency} shows the wall-clock latency of each model across the two toolchains. Vitis outperforms our Calyx-based flow in all cases, particularly on the CNN model. This performance gap is largely due to limitations in Calyx's memory model: Calyx only supports single-dimensional memories, requiring us to flatten multi-dimensional tensors. As a result, access indices—often affine expressions of loop variables—are lowered to hardware as expensive arithmetic operations like multiplication and modulo, which introduce latency. Furthermore, the CNN model's convolution and pooling layers naturally lead to deeper loop nests, amplifying this cost.

Table~\ref{tab:model-resources-all} reports resource utilization across the same models. We observe that Vitis uses more BRAMs for storing weights and biases, while Calyx consumes significantly more LUTs and FFs due to its use of explicit FSMs for control. Overall, Calyx incurs higher resource usage—particularly LUTs—because of its verbose control modeling and lack of aggressive scheduling.



\begin{table}[t]
  \small
  \centering
  \caption{Resource usage across models.}
  \label{tab:model-resources-all}
  \begin{tabular}{lrrrrrr}
    \toprule
    \multirow{2}{*}{Resource} &
      \multicolumn{2}{c}{MHA} &
      \multicolumn{2}{c}{CNN} &
      \multicolumn{2}{c}{FFNN} \\
    \cmidrule(lr){2-3}
    \cmidrule(lr){4-5}
    \cmidrule(lr){6-7}
      & Vitis & Calyx & Vitis & Calyx & Vitis & Calyx \\
    \midrule
      LUTs  &  7846 & 33312 & 3136 & 4574 & 2011 & 3730 \\
      FFs   &  4017 & 5561  & 1815 & 1223 & 1281 & 742  \\
      BRAMs &   194 & 71    &  213 & 43   &  43  & 9    \\
      DSPs  &    19 & 67    &   5  & 14   &   5  & 6    \\
    \bottomrule
  \end{tabular}
\end{table}

\subsection{FFNN Memory Partitioning}
\label{sec:banked}
To further understand the impact of memory banking, we consider the FFNN model. In this study, every memory is partitioned cyclically along each dimension. In Vitis, this is done via \texttt{\#pragma ARRAY\_PARTITION}, while in Calyx the partitioning is implemented via our compiler pass. We compare both latency and resource usage across varying banking factors.

\begin{figure}[t]
  \centering
  \includegraphics[width=0.9\columnwidth]{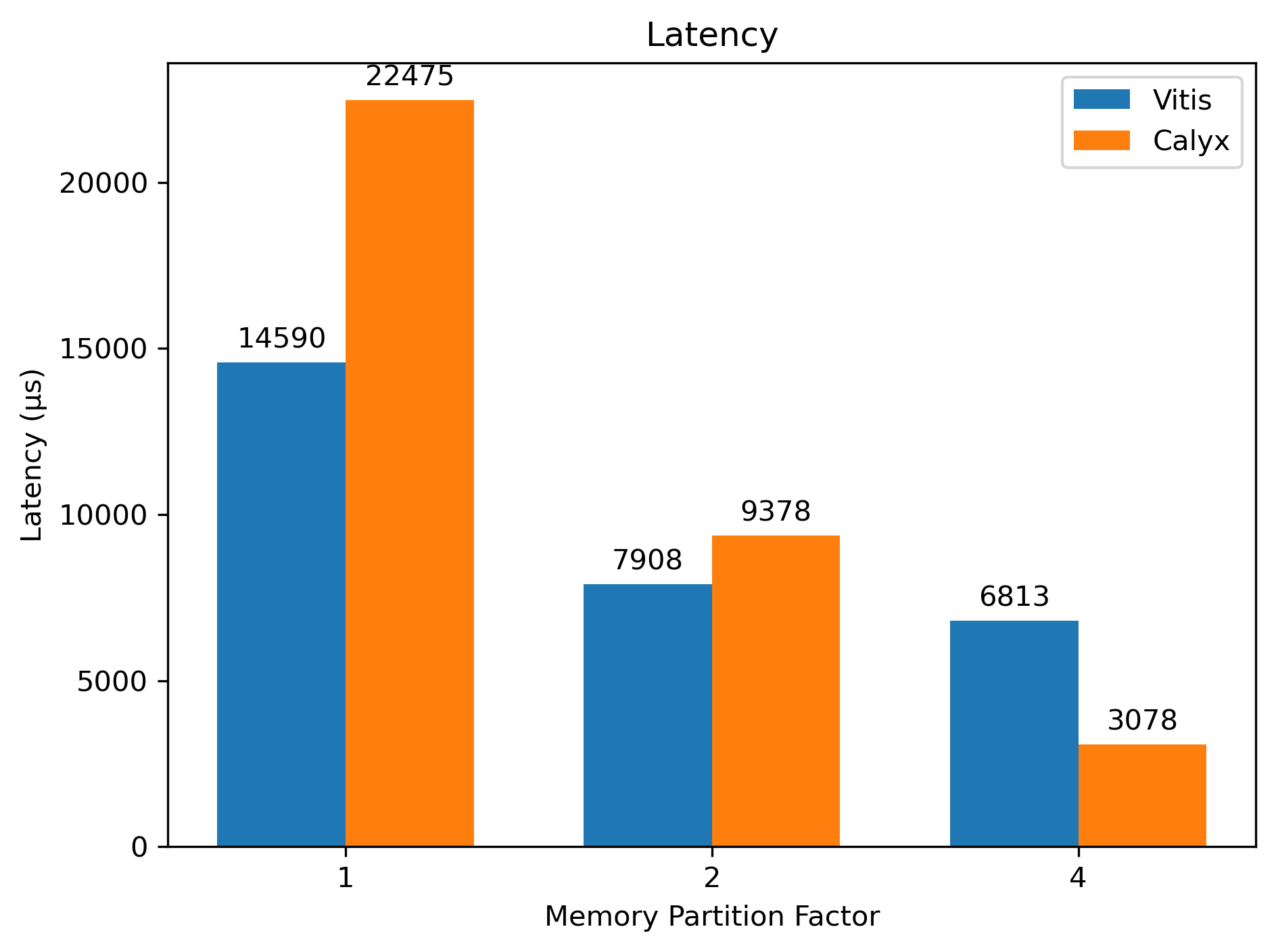}
  \caption{Latency vs. partition factor for FFNN.}
  \label{fig:mem_bank_latency}
\end{figure}

Figure~\ref{fig:mem_bank_latency} shows the latency across different partition factors. For \texttt{factor=1} and \texttt{factor=2}, Vitis performs better. However, at \texttt{factor=4}, Calyx becomes faster. Moreover, the relative speedup for Vitis is modest: increasing the banking factor from 2 to 4 yields diminishing returns, with a speedup of only $7908 / 6813 \approx 1.16$. Given that all matrices are two-dimensional, the theoretical maximum speedup is $2^2 = 4$ under ideal conditions. The limited gain suggests that other bottlenecks remain.

In contrast, the Calyx-based flow shows significant improvement with higher partitioning. The speedup from \texttt{factor=1} to \texttt{factor=2} is $22475 / 9378 \approx 2.40$, and from \texttt{factor=2} to \texttt{factor=4} is $9378 / 3078 \approx 3.05$. This demonstrates the effectiveness of our memory banking analysis and transformation passes, which allow Calyx to unlock more parallelism at the memory level.

Table~\ref{tab:ffnn-resources-all} shows the resource usage corresponding to these experiments. DSP usage is comparable across both toolchains. Vitis uses slightly more BRAMs and FFs, while Calyx consumes significantly more LUTs, particularly at \texttt{factor=4}, due to the additional control logic needed for managing multiple memory banks. These results illustrate the tradeoff between performance and hardware complexity introduced by fine-grained memory partitioning.



\begin{table}[t]
  \small
  \centering
  \caption{Resource usage vs.\ partition factor for the FFNN model}
  \label{tab:ffnn-resources-all}
  \begin{tabular}{lrrrrrr}
    \toprule
    \multirow{2}{*}{Resource} &
      \multicolumn{2}{c}{Partition 1} &
      \multicolumn{2}{c}{Partition 2} &
      \multicolumn{2}{c}{Partition 4} \\
    \cmidrule(lr){2-3}
    \cmidrule(lr){4-5}
    \cmidrule(lr){6-7}
      & Vitis & Calyx & Vitis & Calyx & Vitis & Calyx \\
    \midrule
      LUTs  & 2011 & 3730  & 6021 & 13197 & 13799 & 49121 \\
      FFs   & 1281 & 742   & 4036 & 3145  & 15083 & 10657 \\
      BRAMs &  43  & 9     & 39   & 10    & 64    & 20    \\
      DSPs  &   5  & 6     & 7    & 20    & 80    & 69    \\
    \bottomrule
  \end{tabular}
\end{table}

\section{Conclusion}
This work presents a complete open-source compilation toolchain that transforms PyTorch programs into synthesizable hardware designs using Allo, MLIR, CIRCT, and Calyx. We bridge the gap between software-level ML programs and FPGA-executable hardware accelerators by implementing support for structured control flow, function modeling, floating-point arithmetic, and memory layout transformations.

We focus in particular on optimizing memory access concurrency through static memory banking and control restructuring. We leave hardware pipelining, resource sharing, and compilation of the backward pass for future work. Our evaluation on representative ML models shows that while the Calyx-based flow trails behind commercial tools like Vitis in general-purpose scheduling and resource efficiency, it demonstrates promising performance gains when aggressive memory partitioning is applied. These results highlight the potential of Calyx as a research and prototyping platform for hardware accelerator compilation, especially in settings where open-source and customization are prioritized.


\bibliographystyle{ACM-Reference-Format}
\bibliography{references}

@String{Computing = "Computing" }

@inproceedings{calyx,
author = {Nigam, Rachit and Thomas, Samuel and Li, Zhijing and Sampson, Adrian},
title = {A compiler infrastructure for accelerator generators},
year = {2021},
isbn = {9781450383172},
publisher = {Association for Computing Machinery},
address = {New York, NY, USA},
url = {https://doi.org/10.1145/3445814.3446712},
doi = {10.1145/3445814.3446712},
booktitle = {Proceedings of the 26th ACM International Conference on Architectural Support for Programming Languages and Operating Systems},
pages = {804 - 817},
numpages = {14},
location = {Virtual, USA},
series = {ASPLOS '21}
}

@misc{CIRCT,
  author       = {{LLVM CIRCT. “CIRCT” / Circuit IR Compilers and Tools}},
 url = {https://github.com/llvm/circt},
}

@misc{HardFloat,
    author = {Hauser, John R.},
    title = {Berkeley HardFloat},
    year = {2019},
    url = {https://github.com/ucb-bar/berkeley-hardfloat}
}

@article{Allo,
author = {Chen, Hongzheng and Zhang, Niansong and Xiang, Shaojie and Zeng, Zhichen and Dai, Mengjia and Zhang, Zhiru},
title = {Allo: A Programming Model for Composable Accelerator Design},
year = {2024},
issue_date = {June 2024},
publisher = {Association for Computing Machinery},
address = {New York, NY, USA},
volume = {8},
number = {PLDI},
url = {https://doi.org/10.1145/3656401},
doi = {10.1145/3656401},
journal = {Proc. ACM Program. Lang.},
month = jun,
articleno = {171},
numpages = {28}
}

@article{hls4ml,
    author = "Duarte, Javier and others",
    title = "{Fast inference of deep neural networks in FPGAs for particle physics}",
    eprint = "1804.06913",
    archivePrefix = "arXiv",
    primaryClass = "physics.ins-det",
    reportNumber = "FERMILAB-PUB-18-089-E",
    doi = "10.1088/1748-0221/13/07/P07027",
    journal = "JINST",
    volume = "13",
    number = "07",
    pages = "P07027",
    year = "2018"
}

@INPROCEEDINGS{fpgaconvnet,
  author={Venieris, Stylianos I. and Bouganis, Christos-Savvas},
  booktitle={2016 IEEE 24th Annual International Symposium on Field-Programmable Custom Computing Machines (FCCM)}, 
  title={fpgaConvNet: A Framework for Mapping Convolutional Neural Networks on FPGAs}, 
  year={2016},
  volume={},
  number={},
  pages={40-47},
  doi={10.1109/FCCM.2016.22}}

@inproceedings{DnnWeaver,
author = {Sharma, Hardik and Park, Jongse and Mahajan, Divya and Amaro, Emmanuel and Kim, Joon Kyung and Shao, Chenkai and Mishra, Asit and Esmaeilzadeh, Hadi},
title = {From high-level deep neural models to FPGAs},
year = {2016},
publisher = {IEEE Press},
booktitle = {The 49th Annual IEEE/ACM International Symposium on Microarchitecture},
articleno = {17},
numpages = {12},
location = {Taipei, Taiwan},
series = {MICRO-49}
}

@inproceedings{MLIR,
author = {Lattner, Chris and Amini, Mehdi and Bondhugula, Uday and Cohen, Albert and Davis, Andy and Pienaar, Jacques and Riddle, River and Shpeisman, Tatiana and Vasilache, Nicolas and Zinenko, Oleksandr},
title = {MLIR: scaling compiler infrastructure for domain specific computation},
year = {2021},
isbn = {9781728186139},
publisher = {IEEE Press},
url = {https://doi.org/10.1109/CGO51591.2021.9370308},
doi = {10.1109/CGO51591.2021.9370308},
booktitle = {Proceedings of the 2021 IEEE/ACM International Symposium on Code Generation and Optimization},
pages = {2–14},
numpages = {13},
location = {Virtual Event, Republic of Korea},
series = {CGO '21}
}

@misc{UrbachPetersen22,
  author       = {Mike Urbach and Morten B. Petersen},
  title        = {HLS from PyTorch to SystemVerilog with MLIR and CIRCT},
  howpublished = {Presented at the 2nd Workshop on Languages, Tools, and Techniques for Accelerator Design (LATTE'22)},
  year         = {2022},
  url          = {https://capra.cs.cornell.edu/latte22/paper/2.pdf}
}

@misc{DemmeLandy22,
  author       = {John Demme and Aaron Landy},
  title        = {Using CIRCT for FPGA Physical Design},
  howpublished = {Presented at the 2nd Workshop on Languages, Tools, and Techniques for Accelerator Design (LATTE'22)},
  year         = {2022},
  url          = {https://capra.cs.cornell.edu/latte22/paper/10.pdf}
}

@inproceedings{Dahlia,
author = {Nigam, Rachit and Atapattu, Sachille and Thomas, Samuel and Li, Zhijing and Bauer, Theodore and Ye, Yuwei and Koti, Apurva and Sampson, Adrian and Zhang, Zhiru},
title = {Predictable accelerator design with time-sensitive affine types},
year = {2020},
isbn = {9781450376136},
publisher = {Association for Computing Machinery},
address = {New York, NY, USA},
url = {https://doi.org/10.1145/3385412.3385974},
doi = {10.1145/3385412.3385974},
booktitle = {Proceedings of the 41st ACM SIGPLAN Conference on Programming Language Design and Implementation},
pages = {393–407},
numpages = {15},
location = {London, UK},
series = {PLDI 2020}
}

@inproceedings{ScaleHLS,
author = {Ye, Hanchen and Jun, HyeGang and Jeong, Hyunmin and Neuendorffer, Stephen and Chen, Deming},
title = {ScaleHLS: a scalable high-level synthesis framework with multi-level transformations and optimizations: invited},
year = {2022},
isbn = {9781450391429},
publisher = {Association for Computing Machinery},
address = {New York, NY, USA},
url = {https://doi.org/10.1145/3489517.3530631},
doi = {10.1145/3489517.3530631},
booktitle = {Proceedings of the 59th ACM/IEEE Design Automation Conference},
pages = {1355–1358},
numpages = {4},
location = {San Francisco, California},
series = {DAC '22}
}

\end{document}